\documentclass[preprint]{aastex}
\usepackage{color}
\defcitealias{kim11}{K11}
\newcommand{\paper}{\citetalias{kim11}}
\newcommand{\oops}[1]{#1}
\newcommand{\e}[1]{\underline{#1}}
\newcommand{\vct}[1]{\vec{\mathbf{#1}}}
\newcommand{\het}[1]{\hat{\mathbf{#1}}}
\newcommand{\cs}{c_\mathrm{s}}
\newcommand{\Varm}{V_{\rm arm}}
\newcommand{\rarm}{r_{\rm arm}}
\newcommand{\kmps}{km\,s${}^{-1}$}
\newcommand{\Msun}{$M_\odot$}
\newcommand{\Mspy}{\Msun\,yr${}^{-1}$}
\newcommand{\mach}{\mathcal{M}}
\newcommand{\degree}{\ensuremath{^\circ}}

\shorttitle{AGB WIND INTERACTION WITH SUBSTELLAR COMPANIONS}
\shortauthors{KIM \& Taam}

\begin{document}
\title{Probing Substellar Companions of AGB Stars through Spirals and Arcs}
\author{Hyosun Kim\altaffilmark{1} and Ronald E. Taam\altaffilmark{1,2}}
\altaffiltext{1}{Academia Sinica Institute of Astronomy and Astrophysics, 
  P.O. Box 23-141, Taipei 10617, Taiwan; hkim@asiaa.sinica.edu.tw}
\altaffiltext{2}{Department of Physics and Astronomy, Northwestern University,
  2131 Tech Drive, Evanston, IL 60208; taam@tonic.astro.northwestern.edu}

\begin{abstract}
Recent observations of strikingly well-defined spirals in the circumstellar 
envelopes of asymptotic giant branch (AGB) stars point to the existence of 
binary companions in these objects. In the case of planet or brown dwarf 
mass companions, we investigate the observational properties of the spiral%
-onion shell wakes due to the gravitational interaction of these companions 
with the outflowing circumstellar matter. Three dimensional hydrodynamical 
simulations at high resolution show that the substellar mass objects produce 
detectable signatures, corresponding to density contrasts (10--200\,\%) and 
arm separations (10--400\,AU) at 100\,AU distance, for the wake induced by 
a Jupiter to brown dwarf mass object orbiting a solar mass AGB star. In 
particular, the arm pattern propagates in the radial direction with a speed 
depending on the {\it local} wind speed and sound speed, implying possible 
variations in the arm separation in the wind acceleration region and/or in 
a slow wind with significant temperature variation. The pattern propagation 
speeds of the inner and outer boundaries differ by twice the sound speed, 
leading to the overlap of high density boundaries in slow winds and producing 
a subpattern of the spiral arm feature. Vertically, the wake forms concentric 
arcs with angular sizes anticorrelated to the wind Mach number. We provide 
an empirical formula for the peak density enhancement as a function of the 
mass, orbital distance, and velocity of the object as well as the wind and 
local sound speed. In typical condition of AGB envelopes, the arm-interarm 
density contrast can be greater than 30\,\% of the background density within 
a distance of $\sim10\,(M_p/M_J)$\,AU for the object mass $M_p$ in units 
of Jupiter mass $M_J$. These results suggest that such features may probe 
unseen substellar mass objects embedded in the winds of AGB stars and may 
be useful in planning future high sensitivity/resolution observations with 
ALMA.
\end{abstract}

\keywords{circumstellar matter --- 
  hydrodynamics --- 
  stars: AGB and post-AGB --- 
  stars: late-type --- 
  stars: winds, outflows --- 
  waves}

\section{INTRODUCTION}\label{sec:intr}
In high density environments, substellar companions leave their 
orbital imprints through the gravitational interaction with the 
surrounding circumstellar medium. An example is the observation 
of molecular spiral arms in the protoplanetary disk of AB Aurigae 
\citep{fuk04}, which is considered as evidence of a giant planet 
within the disk \citep{lin06}. It is now well established that 
a spiral wave sets up in the differentially rotating system 
\citep[e.g.,][and references therein]{gol79,mas08} during the 
formation and growth of a planet. This protoplanetary phase lasts 
for a few million years and is followed by a main sequence phase, 
in which the circumstellar disk dissipates and the planet-disk 
interaction ceases.

Planets or brown dwarfs orbiting low and intermediate mass stars 
($M_\ast\le6-8$\,\Msun) may have another opportunity to interact 
with their immediate surroundings before the stars terminate their 
evolution as white dwarf remnants. Stars that have evolved off the 
main sequence, especially on the asymptotic giant branch (AGB), expel 
a large fraction of their mass at rates $\sim10^{-7}-10^{-4}$\,\Mspy, 
forming extensive ($10^{16}-10^{18}$\,cm) cool ($10^2-10^3$\,K) low 
velocity ($3-30$\,\kmps) outflowing envelopes \citep[and references 
therein]{hab03,fon06}. Such intense mass loss rates imply an envelope 
density of $10^{-19}-10^{-16}$ $g$\,cm${}^{-3}$ at 100\,AU distance 
from the stellar center, which is slightly lower than the typical 
density of protoplanetary disks \citep{hay81,aik99}. Similar to the 
protoplanetary case, substellar mass objects that are embedded in the 
dense circumstellar envelopes of evolved giant stars will interact 
with the environment, playing a role in shaping the wind structure 
and also affecting the orbital evolution of the system. 

Recently, the circumstellar envelopes of a few AGB stars have been found 
to possess spiral morphologies (e.g., \citealp{mau06} for AFGL 3068; 
\citealp{tru09} for CIT 6). Such structures cannot be explained by the 
spherically symmetric wind models for isolated stars with either continued 
\citep[e.g.,][]{par58,gil72} or pulsating \citep[e.g.,][]{woo79,wil79,
win00,sim01} mass ejection \citep[for a review, see][]{laf91}, although 
the pulsation models are able to explain the concentric shells and 
arcs of dust detected in a number of post-AGB or proto-planetary phases 
\citep[e.g.,][]{sah98, mau99,kwo01,hri01}. However, direct comparison 
with the detailed observational aspects remains as they likely involve 
the nonlinear process of dust formation in the envelopes. On the other 
hand, the spirals and greatly distended shells/arcs are more naturally 
produced by dynamical effects arising from a binary interaction in the 
expanding wind of the mass losing AGB star \citep{sok94,mas99,he07,edg08}. 
The binary-induced spiral model is supported by the detection of two point 
sources within the amazingly well-defined spiral envelope of AFGL 3068 
\citep{mor06}.

Interestingly, albeit initiated from a different viewpoint, a very 
similar spiral-onion shell structure in three dimensional space was 
described by \citet{kim07}. This study may help facilitate an 
understanding of asymmetric morphologies in circumstellar envelopes 
for cases of companions with low masses, which are not sufficiently 
massive to displace the parent star due to orbital motion. Their 
linear perturbation analysis for gravitationally induced density wakes 
of circularly orbiting objects was motivated by \citeauthor{ost99}'s 
\citeyearpar{ost99} purely analytical formalism of the density wakes 
for linear orbit objects. Follow up studies were carried out for 
binaries \citep{kim08}, for heavy objects showing nonlinear aspects 
\citep{kim09,kwt10}, and for objects in accelerated motion \citep{nam10}. 
\citet{kim09} noticed that the local features of the nonlinear density 
wake are conceptually the same as those of a Bondi-Hoyle-Lyttleton 
accretion column \citep[see][and references therein]{edg04}, and are 
insensitive to the surface condition of the object.

We note that the above theoretical works are based on an initially 
static uniform medium, which differs from the expanding AGB envelopes. 
However, their results deserve to be examined in the extreme limit of 
a slow wind. The focus of these works was on the orbital evolution of 
the perturbing object due to dynamical friction, leaving the details of 
the wake itself in abeyance. In a separate paper \citep[hereafter \paper]
{kim11} the properties of the wake induced by a circularly orbiting object 
at distance $r_p$ with the orbital Mach number $\mach_p$ ($>$1) in a static 
uniform background was investigated. Here, we compare the results for the 
outflowing envelopes to the case of a static background. The resulting 
wake in the latter is characterized by a single-armed Archimedes spiral in 
the orbital plane satisfying $r/r_p=\varphi\mach_p^{-1}+1$\oops{, where 
$\varphi$ is the angular distance along the spiral pattern,} and by arcs in 
meridional planes having the angular size of $2\tan^{-1}(\mach_p^2-1)^{1/2}$. 
Its density contrast is determined by
\begin{equation}\label{equ:aone}
  \alpha_1=\frac{r_B}{|r-r_p|}\frac{1}{(\mach_p^2-1)^{1/2}},
\end{equation}
and the background density modification due to memory effect, where 
$r_B$ is the Bondi accretion radius of the perturbing object. In this 
paper, we investigate the effects of an outflowing background on the 
wake features both analytically and numerically in order to provide 
an interpretative framework of future observations for indirectly 
probing substellar mass objects in AGB envelopes. In \S\ref{sec:simu} 
we describe the setup of the numerical simulation by specifying the 
stellar wind model as a background and the gravitational perturbation 
due to an object orbiting about the central star. In \S\ref{sec:resu} 
we present the numerical results focusing on the shape and morphology 
of the wake as well as its density contrast to the background matter. 
Finally in \S\ref{sec:disc} we discuss our findings with a view toward 
future observations.

\section{SIMULATION SETUP}\label{sec:simu}

\subsection{Background Wind Condition}\label{sec:back}

For a wind characterized by a velocity of $\vct{V_w}$, the density 
distribution $\rho_w$ follows from the steady hydrodynamic conditions
\begin{equation}\label{equ:conw}
  \vct{\nabla}\cdot(\rho_w\vct{V_w})=0,
\end{equation}
and
\begin{equation}\label{equ:momw}
  \vct{V_w}\cdot\vct{\nabla}\vct{V_w}
  = -\frac{\cs^2}{\rho_w}\vct{\nabla}\rho_w-\vct{\nabla}\Phi_\ast^{\rm eff},
\end{equation}
for a background gravitational potential $\Phi_\ast^{\rm eff}$. We 
do not explicitly consider the effects of dust, which is believed 
to play an essential role in driving the wind of cool giant stars, 
but we use an effective potential term $\Phi_\ast^{\rm eff}$ by 
introducing an effective stellar mass $M_\ast^{\rm eff}=M_\ast(1-f)$ 
conceptually to include an additional force due to stellar radiation 
pressure on the wind, $f/r^2$. Here, $\cs$ represents the sound speed 
of the background gas flow, $\cs^2=\gamma p/\rho$ with the adiabatic 
index $\gamma$ in general. In the numerical simulations, we employ 
an isothermal gas approximation ($\gamma=1$). 

\begin{figure}
  \epsscale{0.9}
  \plotone{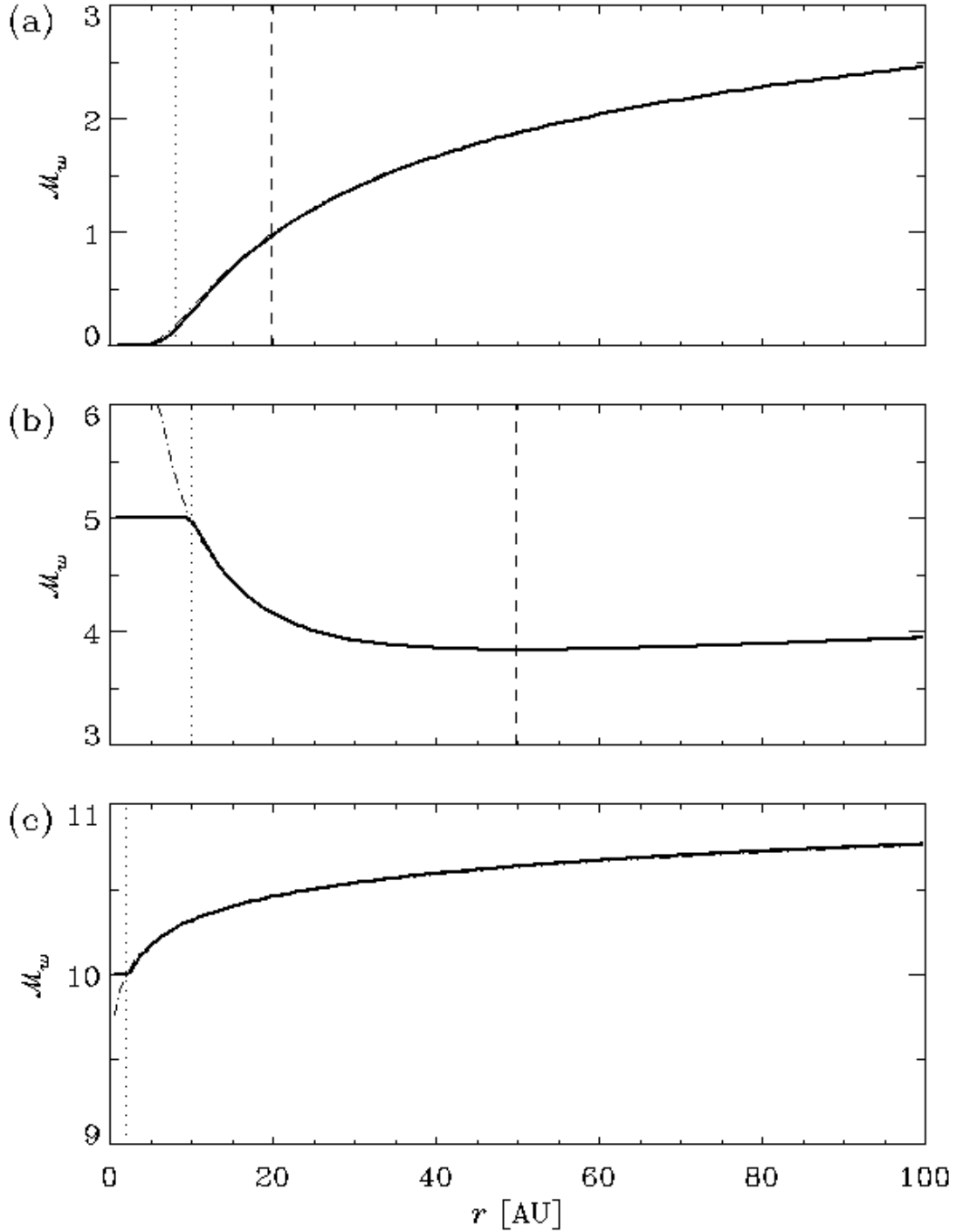}
  \caption{\label{fig:init}
    Background wind Mach number $\mach_w$ as a function of distance $r$ for 
    (a) transonic hydrodynamic wind (model category ``T''), (b) supersonic 
    hydrodynamic wind (model category ``S''), and (c) wind accelerated by 
    additional $r^{-2}$ force (model category ``A''). Thick solid curves 
    show the final stable configuration of the background, following the 
    analytic solution ({\it dot-dashed}) outside the conceptual stellar radius 
    $r_\ast$ ({\it dotted}). Dashed lines in pure hydrodynamic winds indicate 
    the critical radius $r_c=GM_\ast/2\cs^2$ with constant sound speed $\cs$, 
    corresponding to the location of the sonic radius for the transonic branch 
    in (a) and of minimum velocity for supersonic branch in (b).
  }
\end{figure}

In a spherical description, 
the wind material has a density distribution $\rho_w=\rho_w(r)$ 
stratified in the radial direction with a corresponding velocity 
field $\vct{V_w}=V_w(r)\het{r}$ affected by the pressure gradient. 
Given a continuous stellar mass loss rate $\dot{M}_\ast$, mass 
conservation (eq.~[\ref{equ:conw}]) for a radial wind leads to 
$\rho_w(r)=\dot{M}_\ast/4\pi r^2 V_w(r)$. A constant velocity, 
as is observed at large radii, results in the wind attaining 
a $r^{-2}$ density distribution. The details of the velocity 
structure are regulated by the driving mechanism of the wind. 
For instance, a purely gaseous wind without any additional force 
in the momentum equation (eq.~[\ref{equ:momw}]) approaches a 
near-constant velocity regime only at large distances where both 
the stellar gravity and the gas pressure gradient have significantly 
decreased. \citet{par58} derived the isothermal wind solution as 
\begin{equation}\label{equ:park}
  \mach_w^2-\mach_w^2(r_c)-2\ln\left(\frac{\mach_w}{\mach_w(r_c)}\right)
  = 4\ln\left(\frac{r}{r_c}\right)+4\left(\frac{r_c}{r}\right)-4,
\end{equation}
where $\mach_w=V_w/\cs$ represents the wind Mach number and 
the critical radius $r_c=GM_\ast/2\cs^2$ is the so-called sonic 
radius in a transonic wind solution satisfying $\mach_w(r_c)=1$. 
A hydrodynamical wind from the star follows the supersonic 
($\mach_w(r_c)>1$), transonic ($\mach_w(r_c)=1$), or subsonic 
($\mach_w(r_c)<1$) branch of \citeauthor{par58}'s wind solutions 
(Fig.~\ref{fig:init}a--b; \citealp[see also Figure~3.1 of][]{lam99}). 
Any branch of \citeauthor{par58}'s wind solutions shows significant 
variation of the outflowing velocity inside the critical radius on 
the scale of a few tens to thousands AU for typical red giant stars 
(see below for accelerating winds).

The additional outward force due to the stellar radiation pressure 
on dust, which couples with the gas, is commonly regarded as the 
wind driving mechanism of AGB stars \citep{lam99,win00} accelerating
the wind faster so that a near-constant velocity is reached at smaller 
radii. Strong acceleration ($f=1$; $M_\ast^{\rm eff}=0$) modifies 
the isothermal wind to satisfy (see Fig.~\ref{fig:init}c)
\begin{equation}\label{equ:lame}
  \mach_w^2-\mach_w^2(r_\ast)-2\ln\left(\frac{\mach_w}{\mach_w(r_\ast)}\right)
  = 4\ln\left(\frac{r}{r_\ast}\right),
\end{equation}
whose quantities are determined by the conditions at the dust 
condensation radius $r_\ast$. In reality, this is probably not 
a distinct boundary as the wind accelerates gradually and $f$ 
would not reach unity. Our approximations oversimplify the wind 
driving mechanism, but`they are adopted for the background wind 
condition in order to focus on the larger scale structure. 

Although the background wind condition depends on the wind driving 
mechanism, the response of gas to the gravitational perturbation is 
insensitive to the global wind structure, but instead is determined 
by the {\it local} wind quantities. This point will be checked by 
two sets of simulations characterized by the transonic and supersonic 
branches of \citeauthor{par58}'s wind solutions, as well as one set 
of simulations with $f=1$ mimicking strong radiation pressure. Hence, 
it will be confirmed that the wake properties are all determined by 
the wind density $\rho_w(r)$, its expansion speed $V_w(r)$, and the 
sound speed $\cs(r)$ {\it on the spot} in addition to the orbital and 
accretion properties of the perturbing substellar mass object.

\subsection{Model Parameters}\label{sec:para}

In the background wind characterized by its outflowing velocity 
$\vct{V_w}$ and corresponding density $\rho_w$, we introduce a 
perturbation $\delta\vct{V}$ for velocity and $\delta\rho$ for 
density due to the gravitational potential $\Phi_p$ of a substellar 
companion of mass $M_p$ orbiting about the central star at a distance 
$r_p$ with the velocity $V_p$. The response of the gaseous medium is 
governed by the ideal hydrodynamic equations
\begin{equation}\label{equ:cont}
  \frac{\partial\rho}{\partial t}+\vct{\nabla}\cdot(\rho\vct{V})=0,
\end{equation}
and
\begin{equation}\label{equ:mome}
  \frac{\partial\vct{V}}{\partial t}+\vct{V}\cdot\vct{\nabla}\vct{V}
  = -\frac{\cs^2}{\rho}\vct{\nabla}\rho-\vct{\nabla}\Phi_\ast^{\rm eff}
  -\vct{\nabla}\Phi_p,
\end{equation}
for the density $\rho=\rho_w+\delta\rho$ and the velocity fields 
$\vct{V}=\vct{V_w}+\delta\vct{V}$. These equations are integrated 
by the FLASH code of \citet{fry00} in three dimensional Cartesian 
coordinates. We utilize the adaptive mesh refinement (AMR) implementation 
based on the PARAMESH package \citep{mac99}, in order to minimize the 
deviations of the wind from a spherical shape \oops{by assigning} 
sufficient number of grid cells within $r_\ast$ \oops{from the 
center of the mass losing star located at the origin of the 
Cartesian coordinates}. However, to avoid potential artifacts 
due to the rapidly propagating spiral shocks, the refinement 
is set only at the beginning of each run \oops{such that the 
nonuniform subgrids are stationary through out the simulation}. 
\oops{We use the maximum level of refinement up to eight with 
the mother grid number of 64$\times$64$\times$32.}
For the calculation of total gravitational field, the POINTMASS 
implementation is used with small modifications (1) to include two 
sources of gravity, (2) to introduce the gravitational softening radii 
of Plummer-type objects \citep[see, e.g.,][]{bin08}, and (3) to allow 
the motion of the objects.

As the perturber, we consider a substellar mass object with the mass 
$M_p$ greater than 10 $M_J$ where $M_J$ denotes the mass of Jupiter. 
The orbit of the perturber is assumed to be circular, ignoring 
evolutionary processes such as stellar mass loss, mass transport 
to the substellar object, tidal interaction between the star and 
the object, and frictional and gravitational drag forces \citep[see, 
e.g.,][and references therein]{vil09}. The range of orbital radius 
$r_p$ considered in our simulations is 10--150\,AU. We also confirm 
that the effect of the perturber size, defined by its gravitational 
softening radius $r_s$, is minor when its size is sufficiently small, 
exponentially decreasing with decreasing $r_s$ (0.1--10\,AU). 
The key parameters are the velocity of the object, $V_p$, and of the 
wind in the outflowing motion $V_w$, relative to the sound speed $\cs$. 
In the numerical simulations, we treat only a pseudo-isothermal gas 
($\gamma=1.00001$) with the constant sound speed $\cs$ ranging 
between 1--10 \kmps. Given $\cs$, the object speed is arbitrarily 
selected in the range of $\mach_p=0.5$--10, to extend parameter 
space. The wind speeds correspond to a Mach number $\mach_w<11$. 
\oops{We perform 50 models in total, and} Table~\ref{tab:para} 
\oops{lists} the model parameters of our selected sets of simulations 
\oops{and the respective resolutions}. The Bondi accretion radius of 
the perturber, $r_B=GM_p/\cs^2$, is also listed for reference.

The simulation sets are categorized into accelerated (A), supersonic 
(S), and transonic (T) background wind models. The S models are 
further divided according to the wind Mach number into the fast 
($\mach_w=10$) SF and slow ($\mach_w=5$) SS models. For the A models 
that treat the wind accelerated by an arbitrary $r^{-2}$ force, we 
focus on highly supersonic winds, in which the velocity variation is 
significantly less than that of transonic or marginally supersonic winds, 
facilitating analysis of the effects of the other parameters explicitly 
related to the orbiting object. For the S and A models, we start with 
a constant outflow velocity and a $r^{-2}$ density distribution in the 
entire domain, and reset the quantities inside $r_\ast$ every timestep 
to avoid the formal infinite values at the center (eqs.~[\ref{equ:park}] 
and [\ref{equ:lame}]) if an inner boundary is not set. Eventually the 
flow follows the analytic solution, either equation (\ref{equ:park}) 
or (\ref{equ:lame}), determined by the conditions at the reset radius 
$r_\ast$ (Fig.~\ref{fig:init}). The background density $\rho_w$ and 
velocity field $\vct{V_w}$ for the S and A models are obtained from 
the separate runs in which the gravitational influence of the perturbing 
object is suppressed. 

\section{RESULTS}\label{sec:resu}

\begin{figure}
  \plotone{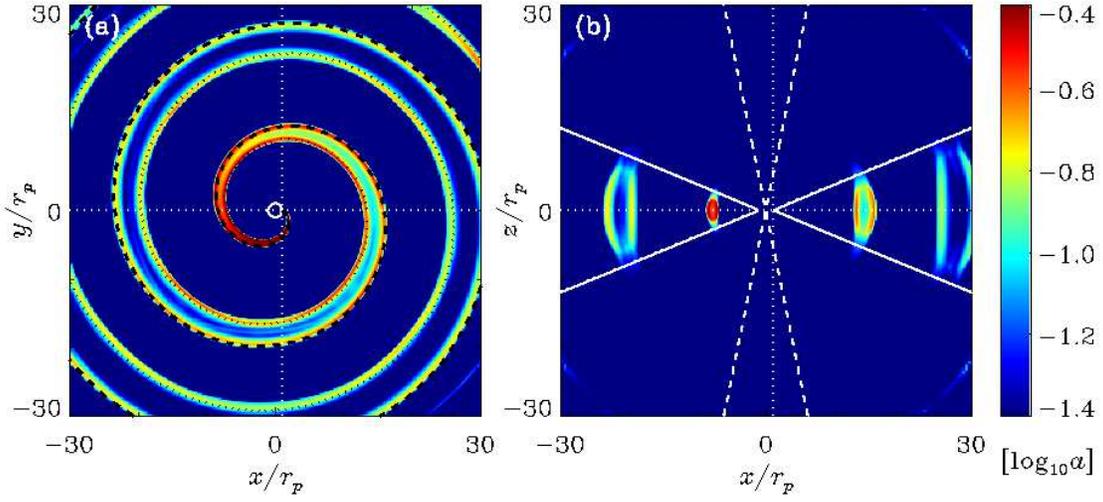}
  \caption{\label{fig:stan}
    Density enhancement $\alpha$ of model with a fast wind (A1) in 
    logarithmic scale. (a) In the orbital plane, the high density 
    region is bounded by quasi-Archimedes spiral arm boundaries, 
    $r/r_p=\varphi\times(\mach_w\pm1)/\mach_p+1$ ({\it black dashed} 
    and {\it dotted} lines) in the polar coordinates ($r$, $\varphi$) 
    with $\varphi$ measured in the clockwise direction. White solid 
    line represents the circular orbit of the perturbing object moving 
    along in the counterclockwise direction and currently located 
    at the intersection of white dotted lines. (b) In a meridional plane, 
    the wake appears to possess arcs with increasing radius along distance 
    from the center. The boundaries are confined by solid lines having 
    angle of $\tan^{-1} [(\mach_p^2-1)^{1/2} (1+0.2\mach_w\mach_p)^{-1}]$ 
    from the orbit. The comparison with the extension limit in 
    the absence of wind (i.e., $\mach_w=0$; {\it dashed}) demonstrates 
    the tendency for the wake to vertically flatten with a fast wind.
    (A color version of this figure is available in the online journal.)
  }
\end{figure}

Figure~\ref{fig:stan} shows the density enhancement factor, 
$\alpha=\delta\rho/\rho_w$, for a fast wind model (A1; 
\oops{see Table~\ref{tab:para}}). Compared 
to the wakes analyzed in \paper\ excluding the background wind, 
the wake in the outflowing environment is radially more extended by 
$\mach_w\pm1$. It is also concentrated to a greater extent toward 
the orbital plane showing a correlation with both wind and object 
Mach numbers, $\mach_w$ and $\mach_p$. In this section, we elucidate 
the details of the shape and morphology of the wake pattern in the 
outflowing circumstance, and determine the arm-interarm density
contrast and its dependence on the wind and object properties.

\subsection{Shape of Spiral Pattern}\label{sec:patt}

\begin{figure}
  \plotone{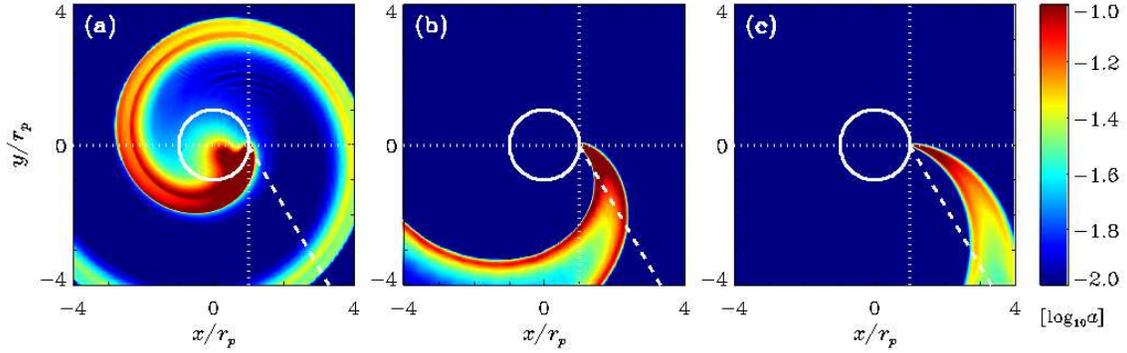}
  \caption{\label{fig:comp}
    Comparison of the shape of spiral pattern between models with the wind Mach
    number $\mach_w$ of (a) 0, (b) 5, and (c) 10. The perturbing object orbits 
    on a circle at distance $r_p$ ({\it solid}, 20 times greater than the Bondi
    accretion radius) in the counterclockwise direction with the orbital Mach 
    number $\mach_p=2$. Its current location is at the intersection of dotted 
    lines. Dashed line denotes the opening angle of the spiral in a static 
    medium, $\Theta=\sin^{-1}\mach_p^{-1}$, from which the corresponding 
    models with faster winds open up of the spiral pattern. The color bar 
    labels the density enhancement $\alpha$ in logarithmic scale. 
    These models are not listed in Table~\ref{tab:para}.
    (A color version of this figure is available in the online journal.)
  }
\end{figure}

The shape of the wake is highly affected by the background flow when 
formed by the gravitational interaction of the orbiting object with 
the background medium. In particular, the wake in the wind of an AGB 
star forms a loosely wound spiral in the orbital plane in contrast to
the case of a static medium. The shape of the spiral is determined by 
the ratio between the pattern propagation speed $\Varm$ and the orbital 
speed of the perturbing object $V_p$ (see \paper),
\begin{equation}\label{equ:patt}
  \frac{d(r/r_p)}{d\varphi}=\frac{\Varm}{V_p},
\end{equation}
which reproduces $\Varm=\cs$ in \citet{kim07} when the wind is absent. In 
Figure~\ref{fig:comp}, we compare the spirals induced by the gravitational 
potential of the orbiting object in different background velocity fields. 
The change of the spiral shapes due to the different wind velocities is
highlighted by the dashed line showing the opening angle of the spiral in 
a static medium. This line matches the spiral shape in the vicinity 
of the object in Figure~\ref{fig:comp}a, while the spirals open further 
with higher wind speeds as displayed in Figure~\ref{fig:comp}b--c. 
This suggests that, in the presence of wind, the propagation speed of 
the spiral pattern in the radial direction, $\Varm$, is faster than 
the sonic speed of the medium. 

For a linear perturbation analysis, we assume that the wind velocity 
$V_w$ is radial and constant in the entire region (see \S\ref{sec:back} 
for the background wind conditions applied in numerical simulations). 
Combining equations (\ref{equ:cont})--(\ref{equ:mome}) in the linearized 
versions for the perturbed velocity $\delta\vct{V}$ under the background 
conditions satisfying equations (\ref{equ:conw})--(\ref{equ:momw}), we obtain 
\begin{equation}\label{equ:comb}
  \frac{D^2 \vct{\delta V}}{D t^2} - \cs^2\ \vct{\nabla}\!\left[\left(
    \vct{\nabla}-\frac{2}{r}\het{r}\right)\cdot\vct{\delta V}\right]
  = -\frac{D}{Dt}\vct{\nabla}\Phi_p
\end{equation}
where $D/Dt=\partial/\partial t+\vct{V_w}\cdot\vct{\nabla}$ is the time 
derivative in the frame moving with the {\em local} background wind. 
Due to the external gravity on the right hand side of the equation, 
the perturbed gas forms a single-armed spiral-like structure, similar 
to one in the static background analyzed in detail by \paper. Using 
spherical harmonics for the perturbation, in general, in the form of 
$\vct{\delta V}=\vct{\xi}(r)P_l^m(\cos\theta)\exp(ikr+im\varphi-i\omega t)$, 
where $\vct{\xi}$ is a slowly varying function of radius, we can write
\begin{equation}
  \left(\begin{array}{lll}
    (Kr)^2-(kr)^2 & i(kr+i)(A+\cot\theta) & -(kr+i)M\\
    iAkr          & (Kr)^2-L^2             & iM(A-\cot\theta)\\
    -Mkr          & iM(A+\cot\theta)       & (Kr)^2-M^2
  \end{array}\right)
  \left(\begin{array}{l}\xi_r\\\xi_\theta\\\xi_\varphi\end{array}\right)
  =\left(\begin{array}{l}0\\0\\0\end{array}\right),
\end{equation}
where $K$, $A$, $M$, and $L^2$ are defined as $K=(\omega-V_wk)/\cs$, 
$A=\frac{d}{d\theta}\ln P_l^m(\cos\theta)$,  $M=m/\sin\theta$, and 
$L^2=l(l+1)+(1-m^2)/\sin^2\theta$, respectively. 
In the WKB approximation ($kr\gg1$ and $Kr\gg1$), the 
dispersion relation yields $\omega=(V_w\pm\cs)\,k$, indicating that
the propagation speed of the pattern is $V_w\pm\cs$ at large 
distances, where $V_w$ and $\cs$ are {\em local} quantities.

\begin{figure}
  \plotone{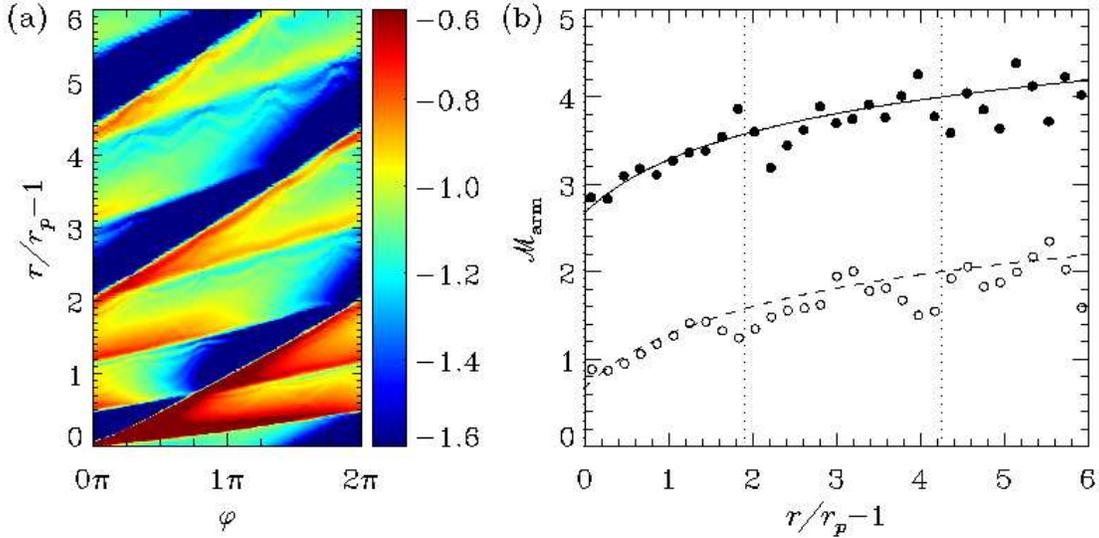}
  \caption{\label{fig:patt}
    Density enhancement of the spiral structure in logarithmic color scale 
    plotted in the polar coordinates ({\it left}), in which the slope of 
    the pattern corresponds to the radial propagation speed of the spiral 
    arm pattern relative to the orbital speed ({\it right}) for T model. 
    Solid and dashed lines represent $\mach_w\pm1$, tracing the slopes of 
    outer ({\it closed circle}) and inner ({\it open circle}) boundaries of
    spiral arm, and the vertical dotted lines denote the turns of outer spiral.
    (A color version of this figure is available in the online journal.)
  }
\end{figure}

In order to check $\Varm$ in the numerical simulations, a density map is 
displayed in Figure~\ref{fig:patt}a for the pattern in polar coordinates 
($\varphi,\,r$) in which the slope of the high density arm boundaries 
represents the velocity ratio $\Varm/V_p$. Numerical differentiation 
along the high density boundary indeed gives the propagation speed of 
$V_w+\cs$ for the outer boundary and $V_w-\cs$ for the inner boundary 
(Fig.~\ref{fig:patt}b), depending on the {\it local} wind properties 
and not on the background wind type. This result implies that in the 
warm central region where the thermal sound speed is relatively large, 
the shape of spiral must take account of the sound speed as well as 
the expansion speed of the envelope. The spacing of the arm is not 
necessarily constant, but varies especially in the wind acceleration 
zone and in slow wind envelopes. In Figure~\ref{fig:patt}a, the outer 
boundary reaches $r/r_p-1=2.0$ and 4.3 after the first and second turns 
(see also vertical dotted lines in Fig.~\ref{fig:patt}b) thus the arm 
spacing increases by 15\% (from $\Delta r/r_p=2.0$ to 2.3). For the 
inner boundary, the locations of $r/r_p-1=0$, 0.5, 1.2, 2.0, 3.0, 4.2, 
and 5.4 indicate $\Delta r/r_p=0.5$, 0.7, 0.8, 1.0, 1.1, and 1.2 for 
the respective turns. That is, the arm spacing changes significantly 
at the first a few turns, where the wind velocity varies considerably
in this model, and the increase rate reaches to 140\% at the sixth
turn relative to the arm spacing at the first turn.

Of particular importance is the overlap of the inner and outer arm 
boundaries seen in Figure~\ref{fig:patt}a at $r_{\rm over}/r_p=1.7$, 3.0, 
and 5.1. In outflowing envelopes the outer and inner boundaries propagate 
at different speeds (by $2\cs$), causing overlaps, unlike the unique 
propagation speed $\cs$ in a static background (\paper). Thus the 
detailed structure of the spiral in the outflowing envelope can be more 
complicated than in a static background. The difference in propagation 
speeds results in the overlap of boundaries at resonant positions 
$r_{\rm over}/r_p$ given by the integration of equation (\ref{equ:patt}) as 
$(V_w+\cs)/V_p\times\varphi$ $\simeq$ $(V_w-\cs)/V_p\times(\varphi+2\pi\times 
integer)$, assuming that the pattern propagation speed does not change 
significantly. Hence, the overlap occurs every $\Delta r_{\rm over}$ 
derived as 
\begin{equation}\label{equ:over}
  \frac{\Delta r_{\rm over}}{r_p}\simeq\frac{\mach_w^2-1}{\mach_p}\ \pi,
\end{equation}
consistent with the intervals between overlays in Figure~\ref{fig:patt}a.

We should also notice the presence of spiral structures in the cases
of subsonic objects (SFx and SSx models). In a static background medium, 
an object with a subsonic orbital speed ($V_p<\cs$) induces the wake 
to be in a smooth distribution rather than confined within a spiral arm 
\citep{kim07}. However, in an outflowing environment, the background wind
sweeps and compresses the induced wake material, resulting in a spiral arm 
shape, as in the cases for supersonic motion.

\subsection{Density Enhancement}\label{sec:jump}

\begin{figure}
  \plotone{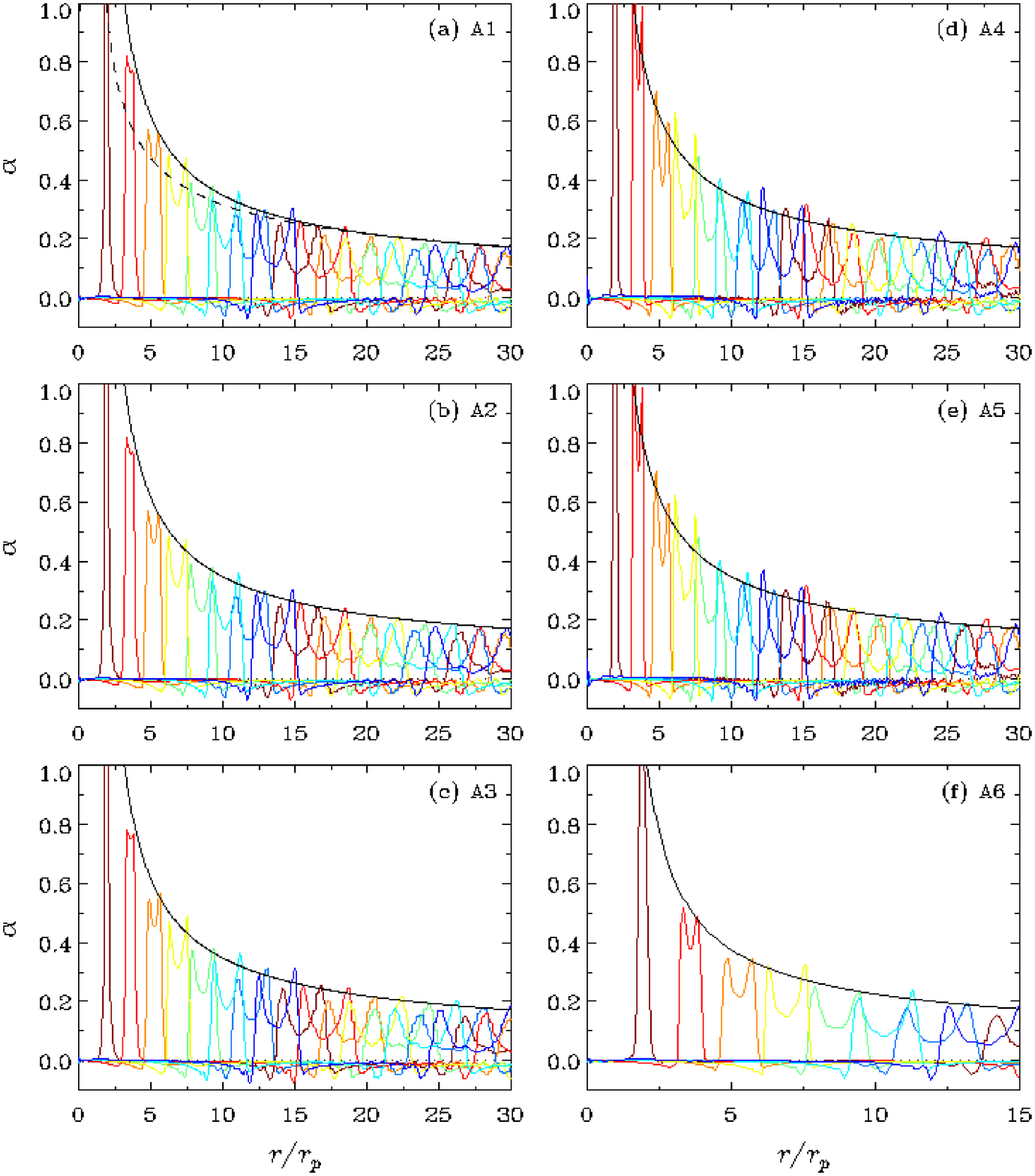}
  \caption{\label{fig:prf1}
    Density enhancement $\alpha$ for models (a) A1, (b) A2, (c) A3, (d) A4, 
    (e) A5, and (f) A6 as a function of the distance normalized by the orbital 
    radius $r_p$ of the perturbing object. The colored lines from red to blue
    display $\alpha$ in the direction apart from the line connecting the object
    and system center with the angle of 0\degree, 45\degree, 90\degree, 
    135\degree, 180\degree, 225\degree, 270\degree, and 315\degree\ trailing 
    the spiral arm. See text for the formula of the complementary black line 
    that outline the peaks.
  }
\end{figure}

Figure~\ref{fig:prf1}a exhibits the density enhancement factor for model A1. 
In order to trace the full structure of the spiral, we plot the profiles 
along eight different directions in the orbital plane with different 
colored lines. One noticeable difference from the corresponding profiles 
in a static background is the level of the zero baseline of the perturbed 
density. Without a wind, the wake always has minimum density with the 
central value of $\alpha=r_B/r_p$ forming a hydrodynamic equilibrium 
(see \paper\ for details). However, in the outflowing medium, this global 
increase of background density is suppressed since the material is swept 
out of the central part by the wind before its pressure balances the 
gravitational potential at the central part. This wind sweeping effect also 
precludes the wake from developing the nonlinear features described by 
\citet{kwt10}. In Figure~\ref{fig:prf1}a, the peak values of the density 
enhancement decrease with distance as outlined by a solid line (see below). 
This is approximated by a dashed line, $\alpha=(r_B|r-r_p|^{-1})^{1/2}$, 
with the power law index of 0.5 except for the steeper profile in the 
innermost part. 

In order to explore the dependence of the density enhancement $\alpha$ on 
the properties of object and envelope, we compare in Figure~\ref{fig:prf1} 
the profiles of $\alpha$ along distances in the orbital plane for 
the A models. Compared to the fiducial model (A1), we investigate 
the effects of the stellar mass loss rate $\dot{M_\ast}$ (A2), the 
size of the star $r_\ast$ defined by the numerical reset radius 
(A3), the gravitational softening radius of the perturbing object 
$r_s$ (A4), the mass of the object $M_p$ as well as the sound speed 
of the background gaseous medium $\cs$ (A5), and the orbital radius 
of the object $r_p$ (A6). By comparing model A1 with models A2 and A3, 
it is found that the perturbed density $\alpha=\delta\rho/\rho_w$ does 
not depend on the mass loss rate $\dot{M_\ast}$ and the radius $r_\ast$ 
at which the mass loss occurs, while the overall density structure 
$\rho_w(r)=\dot{M}_\ast/4\pi r^2 V_w(r)$ is regulated by the mass 
loss rate $\dot{M_\ast}$ and the background wind velocity $V_w(r)$. 
The independence of density enhancement on the mass loss rate is a 
natural consequence for an isothermal gas. Reducing the gravitational 
softening radius $r_s$ of the perturbing object by an order of magnitude
from $0.1\,r_B$ (A1) to $0.01\,r_B$ (A4) sharpens the density profiles 
to some extent but insignificantly changes the overall features. The 
identity of density profiles in the models A4 and A5 indicates that 
the density enhancement $\alpha$ depends merely on the Bondi accretion 
radius $r_B\propto M_p/\cs^2$, and not on the object mass $M_p$ or the sound 
speed $\cs$ individually; it is also checked that $\alpha$ is proportional 
to $M_p$ and $\cs^{-2}$ using models A5-1 and A5-2, respectively. For model 
A6, the perturbing object is located at a distance twice as large from the 
central star than the orbital distance in the other A models, resulting in 
a similar value of $\alpha$ at large distances $r \gg r_p$ although the 
width and interval of the spiral have increased with the larger orbital 
radius $r_p$.
 
\begin{figure}
  \plotone{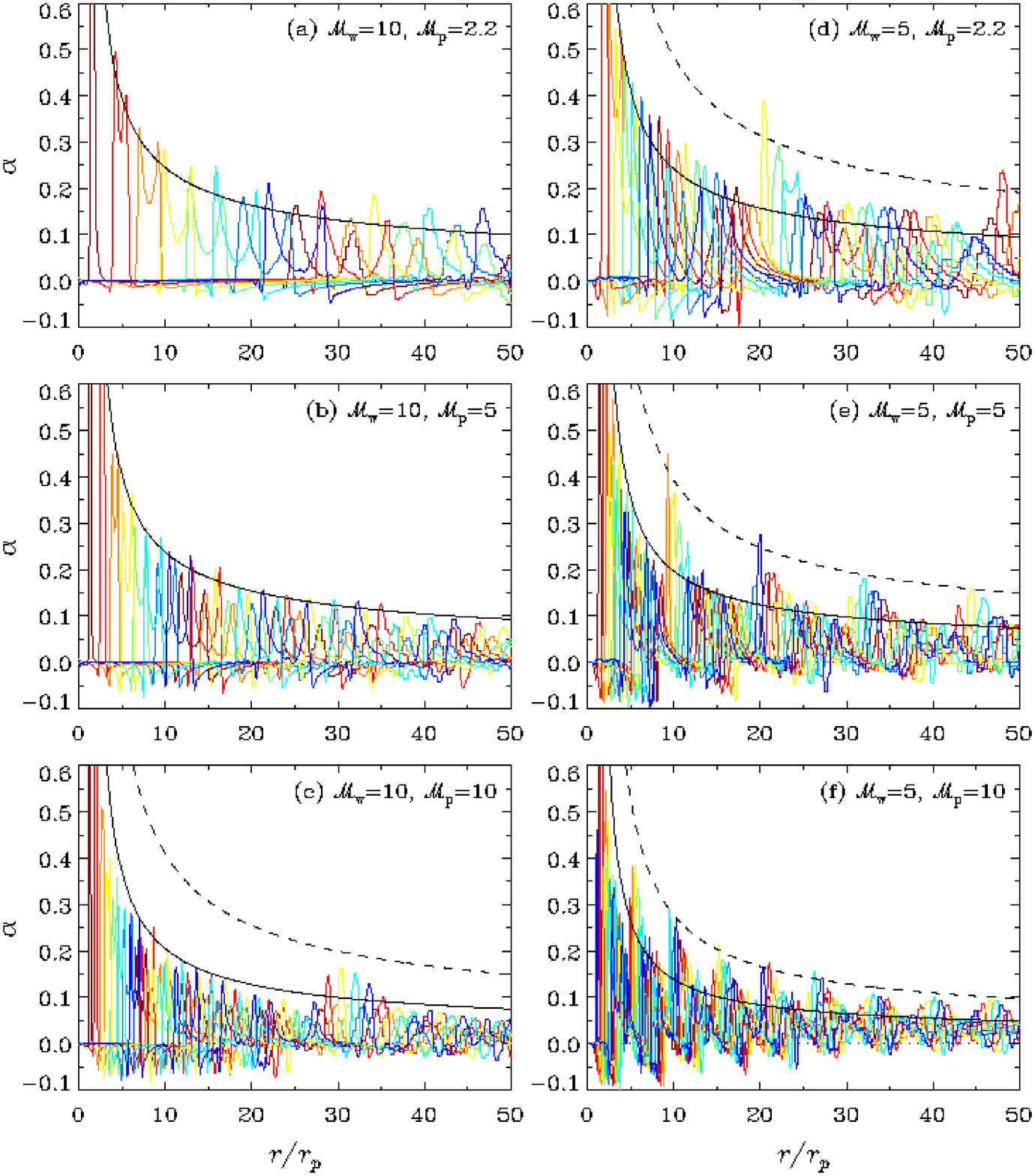}
  \caption{\label{fig:prf2}
    Same as Fig.~\ref{fig:prf1} except for models (a) SFs, (b) SFm, (c) SFf, 
    (d) SSs, (e) SSm, and (f) SSf. The subpattern showing high density groups 
    of peaks (outlined by {\it dashed} line) is due to the overlap between 
    the inner and outer boundaries of the spiral wake pattern. 
  }
\end{figure}

In Figure~\ref{fig:prf2}, we further examine the physical parameters that 
affect the wake density using the S models, focusing on the velocities 
related to the object motion ($V_p$) and the wind expansion ($V_w$) in 
units of background sound speed $\cs$. Comparison between SF models 
(Fig.~\ref{fig:prf2}a--c) and between SS models (Fig.~\ref{fig:prf2}d--f) 
separately reveals a definite anticorrelation of the pattern spacing with 
the orbital Mach number $\mach_p$ given a nearly constant wind Mach 
number $\mach_w$ (9.5--10 in SF; 3.9--5.2 in SS). On the contrary, one can 
also find the decrease of the pattern spacing with a smaller wind Mach 
number $\mach_w$ by comparing the profiles in (a) and (d), (b) and (e), 
or (c) and (f). The reduced spacing of the spiral arm pattern facilitates 
overlaps between the inner and outer arm boundaries characterized by 
different propagation speeds with respect to each other. As a consequence 
of the overlaps, the wake develops subpatterns showing additional increases 
in peak densities at the overlap positions, $r_{\rm over}$, with gradual 
decreases between the overlap positions. A rough estimate for the 
overlap intervals, $\Delta r_{\rm over}/r_p$ from equation (\ref{equ:over}),
with the assumption of constant $\mach_w$ is consistent with the intervals 
between the highest density peaks in Figure~\ref{fig:prf2}: $>50$ for (a) 
and (b), $\sim30$ for (c), $20-30$ for (d), $10-15$ for (e), and $5-8$ for 
(f). The peak densities in Figure~\ref{fig:prf2} (excluding the extra peaks 
due to the subpattern) tend to increase with lower orbital Mach number and 
higher wind Mach number at $r\gg r_B$, albeit the dependence on the 
wind Mach number is not as clear. 

We generalize all the parameter dependence described above, using 50 models
of A-, S-, and T-type background winds in the range of $\mach_w\leq11$ 
with $\cs=1$--10 \kmps\ for perturbing objects of $\mach_p=0.5$--10 and 
$r_B/r_p=0.05$--1. By modifying equation (\ref{equ:aone}), based on \paper's 
analysis for the simpler case of a static uniform background, we obtain 
an empirical formula for the minimum value of the density peaks (black solid 
lines in Fig.~\ref{fig:prf1} and \ref{fig:prf2}) described as follows:
\begin{equation}\label{equ:jump}
  \alpha_{peak}\gtrsim\frac{r_B}{|r-r_p|}\left(\frac{
    \frac{|r-r_p|}{r_B}\mach_w^2+10\mach_p\mach_w+1}{\mach_w^2+|\mach_p^2-1|}
  \right)^{1/2} \qquad {\rm for}\ \ \mach_w>1,
\end{equation}
which yields $\alpha_1$ (eq.~[\ref{equ:aone}]) when $\mach_w=0$. 
We note that this equation is suitable for supersonic winds since the 
winds in our simulations are subsonic only in the innermost region in 
T-type models, which does not sufficiently contribute to the overall 
profile included in our analysis. A subsonic wind likely leads to a 
more complicated wake description because (1) it is not able to fully 
sweep off the delayed perturbations from the object in the past orbits, 
and (2) concurrently it could cause nonlinear effects, in addition to 
(3) the wind effect, described here, making the inner and outer arm 
boundaries overlap because of the different propagating speeds. The 
dashed lines in Figure~\ref{fig:prf2} outlining the maximum density 
peaks including substructures, $\alpha=2\alpha_{peak}$ is adopted.

\subsection{Vertical Flatness}\label{sec:flat}

As seen in Figure~\ref{fig:stan}b, the vertical structure of a gravitational
wake has a shape of concentric arcs, confined by solid lines representing 
the empirical formula for the vertical extension limit. The angular size of 
the arcs is best fit to the relation
\begin{equation}\label{equ:flat}
  \theta_{\rm arc}
  =2\tan^{-1}\left(\frac{(\mach_p^2-1)^{1/2}}{1+0.2\mach_w\mach_p}\right),
\end{equation}
which is smaller (or further concentrated toward the orbital plane) 
in comparison to its counterpart in a static medium ($\mach_w=0$, dashed; 
see \paper\ for details). 

\begin{figure}
  \plotone{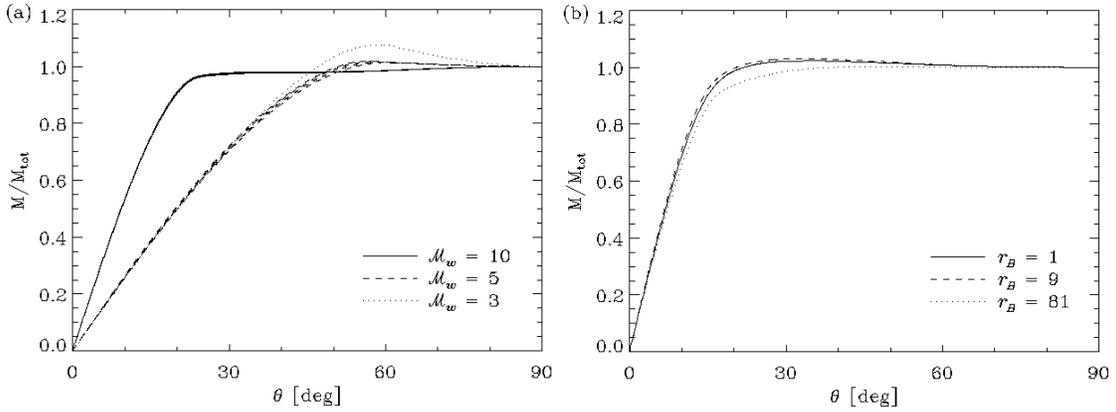}
  \caption{\label{fig:flat}
    Integrated mass $M$ of perturbed density $\alpha$ as a function of the 
    angle from the orbital plane, normalized by the total mass enhancement 
    $M_{\rm tot}$ in the simulation domain. 
    (a) The integrated mass of wake increases gradually and levels off at 
    20\degree\ for $\mach_w\simeq10$ cases ({\it solid}, SF models), but 
    it reaches saturation at 50\degree\ for $\mach_w\simeq3-5$ cases 
    ({\it dashed}, SS models; {\it dotted}, T model). 
    (b) Comparison of the integrated wake mass in models A1 ({\it dashed}), 
    A5-1 ({\it solid}), and A5-2 ({\it dotted}) indicates its weak dependence 
    on the accretion radius $r_B=GM_p/\cs^2$, and moreover on the perturber 
    mass $M_p$ and the sound speed $\cs$ of the gas flow, separately.
  }
\end{figure}

In order to quantify the wake mass distribution in the vertical direction, 
the perturbed density $\alpha$ is integrated as a function of latitudinal 
angle. Figure~\ref{fig:flat}a displays the integrated wake mass normalized 
by the total wake mass for different wind Mach numbers $\mach_w$. 
The dotted and dashed lines \oops{(T and SS models)} show 
gradual increases until reaching unity at $\sim$~50\degree\ for 
$\mach_w\simeq3-5$, while the solid lines \oops{(SF models)} reveal 
that the saturation is already achieved at $\sim$~20\degree\ for a higher 
wind Mach number $\mach_w\simeq10$. Beyond these latitudinal angles, the 
AGB envelopes remain unchanged under the gravitational perturbation of 
the orbiting substellar object. The decrease of the critical angle with 
$\mach_w$ implies a more flattened morphology of the circumstellar matter 
in a faster outflow. This is due to the fact that in fast background winds, 
the secondary's gravitational potential relative to the wind kinematics 
is low so that its perturbation affects only a limited area. In slow 
winds, however, the dependence of the vertical extent on $\mach_w$ is 
weak, as revealed in the resemblance of the dotted ($\mach_w=3$) and 
dashed ($\mach_w=5$) lines in Figure~\ref{fig:flat}a.

The distribution of the integrated wake mass is relatively independent 
of $\mach_p$, $\cs$, $M_p$, and the background wind model. For instance, 
Figure~\ref{fig:flat}a shows the overlap of 4 solid lines \oops{(SFf, 
SFm, SFs, and SFx models)} and 4 dashed lines \oops{(SSf, SSm, SSs, 
and SSx models)}, respectively, for $\mach_p=0.5$, 2.2, 5.0, and 10.0. 
The independence on the perturber mass $M_p$ and the fluid sound speed 
$\cs$ (or the Bondi accretion radius $r_B$ by association) is shown in 
Figure~\ref{fig:flat}b \oops{(A1, A5-1, and A5-2 models)}. Moreover, the 
independence on the wind model is found from similarities between the 
solid lines in Figure~\ref{fig:flat}a (S \oops{models}) and the lines in 
Figure~\ref{fig:flat}b (A \oops{models}) for fast winds ($\mach_w\simeq10$) 
and between the dotted (T \oops{model}) and dashed (S \oops{models}) 
lines in Figure~\ref{fig:flat}a for slower winds ($\mach_w\simeq3-5$). 
In conclusion, the vertical distribution of a wake in an outflowing 
background is determined by the Mach number of the average expansion 
velocity, $\mach_w$ ($\geq1$), only weakly depending on the accretion 
and orbital properties of the perturbing object.

\section{DISCUSSION}\label{sec:disc}

In order to probe the existence of planets and/or brown dwarfs around 
evolved giant stars, we investigate the properties of the gravitational 
wake induced by a substellar mass object in an outflowing environment, 
through hydrodynamic simulations using the FLASH adaptive mesh refinement 
code. The perturbing object of substellar mass $M_p$ is assumed to be 
in circular motion at distance $r_p$ with the orbital speed of $V_p$.
For an isothermal sound speed $\cs$, the expansion speed $V_w(r)$ 
of the background wind is determined by the hydrodynamic condition 
at a reset radius representing the wind-generating location. 
As functions of the object and wind parameters, $M_p$, $r_p$, $V_p$, 
$V_w$, and $\cs$, we quantify the observable properties of the wake, 
revealing that the shape of the spiral and arcs are dependent on viewing 
direction and the density contrast of the structure. 
Conversely, possible future observations of circumstellar spirals 
and arcs will allow us to place constraints on the properties of 
substellar objects in the winds of AGB stars.

We find that the overdense arm pattern of the wake propagates outward 
with the speed of $V_{\rm arm}=V_w\pm\cs$, where the upper and lower signs 
represent the case for the outer and inner boundaries of the arm pattern, 
respectively. Here, the speeds are local, implying that the arm spacing 
varies in the wind acceleration zone and/or in regions where the sound 
speed varies. In order to measure the variation of arm spacing due to the 
wind acceleration, which is believed to occur within up to a few tens of 
stellar radii \citep{hab03}, observations with high angular resolution
($\lesssim$ a few $0.\!\!\arcsec1$ for the closest AGB stars) are required.
Outside the acceleration zone, where the variation of pattern propagation 
speed is negligible, these pattern propagation speeds yield the wake shape 
in the orbital plane, 
\begin{equation}\label{equ:spir}
  r/r_p=\varphi\times(V_w\pm\cs)/V_p+1
\end{equation}
for the outer 
and inner boundaries, respectively, constituting Archimedes spiral shapes. 
If we define the arm spacing $\Delta\rarm$ as the interval of the outer or 
inner boundary individually, the spiral shapes of boundaries simply yield
\begin{equation}\label{equ:del1}
  \Delta\rarm=(V_w\pm\cs)\times\frac{2\pi r_p}{V_p}
\end{equation}
assuming that the pattern propagation speed does not significantly vary 
within one turn. One can estimate the orbital period, $2\pi r_p/V_p$, by 
measuring the observed arm spacing in addition to 
the wind and sound speeds. Alternatively, this is written by
\begin{equation}\label{equ:del2}
  \Delta\rarm=\frac{2\pi(V_w\pm\cs)}{(GM_\ast)^{1/2}}\times r_p^{3/2},
\end{equation}
from which we can predict the orbital distance $r_p$ of the substellar
object in the case that the stellar mass $M_\ast$ is known. We note 
that the sound speed should be taken into account especially in a slow 
wind ($V_w\sim\cs$).

The outer arm boundary propagates at a higher speed than the inner boundary by 
$2\cs$, leading to the broadening of the arm as a function of radial distance. 
Thus, these high density boundaries overlap each other and are in resonance 
at the overlap positions, $r_{\rm over}$. Assuming a constant wind and 
sonic speeds, the spacings of these resonant positions is derived to 
be $\Delta r_{\rm over}/\Delta\rarm=0.5\,(V_w/\cs\mp1)$ (see eqs.\,%
[\ref{equ:over}] and [\ref{equ:del1}]). In cool AGB envelopes, the 
wind is usually much faster than the sound speed ($V_w\gg\cs$), making 
$\Delta r_{\rm over}$ relatively large. For instance, with $V_w=10\cs$, 
which is easily found for AGB stars, the outer and inner arm boundaries 
can meet only after five turns. For the case of an observational detection 
of only parts of the spiral, especially when the partial spiral is expected 
over the distance of $\Delta r_{\rm over}$, a more careful analysis is 
required to avoid misidentifying the outer and inner boundaries. 

We have also provided an empirical formula (eq.~[\ref{equ:jump}]) for 
the arm-interarm density contrast, $\alpha=\delta\rho/\rho_w$, along 
a normalized distance $r/r_p$ as a function of $V_w/\cs$, $V_p/\cs$, and 
$r_B/r_p=GM_p/(\cs^2r_p)$. Using this empirical formula, we estimate the 
properties of a Jupiter wake in the stellar wind when our Sun becomes a giant 
of size 1\,AU ($M_\ast=0.8$\,\Msun, $\dot{M_\ast}=2\times10^{-7}$\,\Mspy, 
$T_\ast=3000K$; according to \citealp{hur00}). The wind speed $V_w$ is 
set to be $\sim$\,10\,\kmps\ based on the trend between the mass-loss 
rate and the envelope expansion speed \citep[see Fig.~16 in][]{fon06}, and 
the sonic speed $\cs$ is assumed to be 1\,\kmps. The estimated orbital 
speed $V_p=(GM_\ast/r_p)^{1/2}$ of Jupiter in situ ($r_p=5$\,AU) is 
12\,\kmps, corresponding to the arm-interarm density contrast of only
7\,\% of the background density at distance of 100\,AU. The density 
contrast does not significantly depend on the orbital distance, i.e., 
6--11\,\% for $r_p=3$--30\,AU. Thus, it is difficult to 
detect the gravitational wake of a Jupiter mass object with the current 
observational limitations of sensitivity and angular resolution. 
In the same range of orbital distance, a 10 Jupiter mass object can create 
a gravitational wake with the density contrast of 30--44\,\% at 100\,AU; 
and for a brown dwarf mass (0.07\,\Msun), the contrast increases 
to 160--220\,\%, which may be detectable with the high sensitivity 
performance of the Atacama Large Millimeter/submillimeter Array (ALMA). 
We note that these numerical values are lower limits since the peak 
density contrast at the arm boundary can be higher with a realistic 
size of the object much smaller than $\geq0.1$\,AU employed in this 
study; but the effect of the object size $r_s$ is not significant 
unless the size is comparable to the accretion radius $r_B$. The 
required spatial resolution to distinguish the arm pattern separation 
is 12--370\,AU depending on the orbital distance ($r_p=3$--30\,AU). 
On a larger scale, a distance corresponding to 5 times this arm 
separation (in this case of $V_w/\cs=10$) is the distance between 
overlaps showing a higher density contrast by a factor of 2. 

The case for fast winds compared to the sonic and object speeds is the 
most relevant case for many observed AGB envelopes ($\gtrsim$~10\,\kmps). 
For a substellar object orbiting at a distance, 
$r_p>10\,(M_\ast/M_\odot)\,(V_w/10\,\textrm{km\,s}{}^{-1})^{-2}\,{\rm AU}$, 
the density contrast 
at large distance ($r\gg r_p,\,r_B$) can be simply written as
\begin{equation}\label{equ:fast}
  \alpha_{peak}\gtrsim\left(\frac{r_B}{r}\right)^{1/2}
  \qquad {\rm for}\ \  V_w>V_p\gg\cs,
\end{equation}
from which the measurement of the density contrast provides the value 
of the Bondi accretion radius $r_B=GM_p/\cs^2$ (or the object mass $M_p$ 
with a given sound speed) for the structure at a known distance from 
the AGB star. Under the fast wind assumption, the density contrast 
shows a shallower decrease with distance, compared to its counterpart 
in an initially static background decreasing as $r^{-1}$ (\paper). 
This shallow decrease in the density contrast facilitates observations 
at great distances from the star, which is favorable to avoid the high 
obscuration due to either the bright stellar glare or the dense central 
environment. Equation (\ref{equ:fast}) indicates that the density contrast 
between arm and interarm regions can be greater than $\alpha_{peak}$ for 
distances where 
\begin{equation}\label{equ:}
  \left(\frac{r}{\rm AU}\right)\,\gtrsim\,\left(\frac{M_p}{M_J}\right)
  \left(\frac{\cs}{{\rm km\,s}^{-1}}\right)^{-2} {\alpha_{peak}}^{-2}.
\end{equation}
Assuming that we can distinguish a density fluctuation of 30\,\% 
($\alpha_{peak}\sim0.3$), the wake due to an object of mass $M_p$ 
would be observable in the central area within a distance of 
$\sim10\,(M_p/M_J)$\,AU in the case of $\cs=1$\,\kmps. 
That is, we may expect to observe the wake of a $10M_J$ mass planet 
within 100\,AU with a sensitivity corresponding to a density contrast 
of 30\,\%. Alternatively, it may be necessary to take account 
of the density enhancement due to the planetary wakes when modeling the 
overall profile of circumstellar envelopes using observational data in 
which the wakes are not resolved.

In the regime for which the background wind is slower than the orbital velocity
($V_p\gtrsim V_w\gg\cs$), corresponding to an object in a close orbit at 
$r_p\lesssim10\,(M_\ast/M_\odot)\,(V_w/10\,{\rm km\,s^{-1}})^{-2}\,{\rm AU}$,
the peak density contrast at large distance ($r\gg r_p,\,r_B$) is roughly 
proportional to the velocity ratio $V_w/V_p$. In this regime, the effect 
of the wind outflow in opening the spiral pattern competes with the object 
orbital motion tightening the pattern (see eq.~[\ref{equ:spir}]), causing 
a higher density jump at the more normal shock boundary to the instantaneous 
flow.

\oops{Although the analysis in this paper only treats a substellar 
mass object orbiting about the AGB star fixed at the system center, 
our results can be extrapolated to the gravitational density wake
of the companion in a stellar binary system. For instance in an AGB 
envelope, characterized by a temperature of 100\,K ($\cs=1$\,\kmps), 
a solar mass companion would produce a density fluctuation of a factor 
of 1.6 to 1.3 ($\alpha_{peak}\sim0.6$--$0.3$ from eq.~[\ref{equ:fast}]) 
at distance of 3,000--10,000\,AU from the center of the mass losing 
star, where the spiral pattern is found in AFGL 3068 \citep{mau06}. 
The density fluctuation from the companion's wake is smaller than 
that estimated from the scattered light observation (the factor of 
up to $\sim5$), indicating that the spiral pattern of AFGL 3068 is 
likely caused by the reflex motion of the AGB star as suggested in 
\citet{mau06}, following earlier theoretical work, for example, by 
\citet{sok94}. However, in the central 1\arcsec\ region ($<$ 1,000\,AU), 
where the density fluctuation due to the companion is larger, ALMA 
would possibly resolve a double spiral structure due to the individual 
stars. To compare the contribution of the two different mechanisms, 
we plan to study the density enhancement due to the reflex motion 
of the mass losing star, similar to the study here. The shapes of 
two spirals (e.g., pitch angle, arm spacing) are expected to be same, 
because the spiral shape is determined by the local wind properties, 
$V_w$ and $\cs$. But the spiral of the AGB star is probably detached 
from the star because of the stellar outflow, contrary to the spiral 
of the companion attached to the object.}

Our work suggests directions for future research. For example, a detailed 
study of the inner regions of the AGB envelope is necessary to examine the 
effects of gas heating/cooling, dust anisotropic distribution, and possibly 
stellar pulsation on the properties of the spiral arm pattern. In addition, 
investigations of binary systems over a wider range of mass ratios should 
be examined to explore the modifications in the spiral/arc patterns in the 
regime where the effect of the reflex motion of the AGB star is also important.
Finally, radiative transfer modeling of the molecular line emission and dust 
continuum are encouraged in order to compare theoretical models with observed 
structures in detail.

\acknowledgments
\oops{We acknowledge a helpful report from an anonymous referee.} H.K is 
grateful to Francisca Kemper and Kanak Saha for fruitful comments through 
reading the manuscript, Paul M.\ Ricker for advice on computational issues, 
and Noam Soker for discussion on vertical extension.
This research is supported by the Theoretical Institute for Advanced 
Research in Astrophysics (TIARA) in the Academia Sinica Institute of 
Astronomy and Astrophysics (ASIAA). The computations presented here 
have been performed through the ASIAA/TIARA computing resource, using 
FLASH3.0 code developed by the DOE-supported ASC/Alliance Center for 
Astrophysical Thermonuclear Flashes at the University of Chicago.

\clearpage

\begin{deluxetable}{lrrrrrrrrrcrrrc}
\tablecolumns{14}
\tabletypesize{\scriptsize}
\tablewidth{0pc}
\tablecaption{Parameters for numerical simulations\label{tab:para}}

\tablehead{\colhead{} & \multicolumn{5}{c}{Object Properties} & \colhead{}
 & \multicolumn{4}{c}{Background Properties} & \colhead{}
 & \multicolumn{3}{c}{Domain \oops{Information}} \\[0.1pt]
\cline{2-6} \cline{8-11} \cline{13-15} \\[-2ex]
\colhead{Model} & \colhead{$M_p$}
 & \colhead{$r_B$} & \colhead{$r_s$} & \colhead{$r_p$} & \colhead{$\mach_p$}
 & \colhead{} & \colhead{$\dot{M}_\ast$} & \colhead{$r_\ast$}
 & \colhead{$\overline{\mach_w}$ ($\mach_{w,p}$)} & \colhead{$\cs$}
 & \colhead{} & \colhead{$\oops{L}$} & \colhead{$\Delta \oops{L}$} 
 & \colhead{\oops{refinement}} \\
\colhead{} & \colhead{[\Msun]}
 & \colhead{[AU]} & \colhead{[AU]} & \colhead{[AU]} & \colhead{}
 & \colhead{} & \colhead{[\Mspy]} & \colhead{[AU]}
 & \colhead{} & \colhead{[\kmps]}
 & \colhead{} & \colhead{[AU]} & \colhead{[AU]} & \colhead{}}
\startdata

\sidehead{Accelerated Wind via $r^{-2}$ Force ($M_\ast^{\rm eff}=0$~\Msun)}
\hline\\[-2ex]
A1 & 0.01 & 9 & 1.0 & 10 & 5.0 && $1\times10^{-6}$ & 2 & 11 (10.3) & 1
 && 300 & 0.15 -- 2.34 & \oops{7 -- 3} \\ 
A2 & 0.01 & 9 & 1.0 & 10 & 5.0 && \e{$1\times10^{-4}$} & 2 & 11 (10.3) & 1
 && 300 & 0.15 -- 2.34 & \oops{7 -- 3}\\ 
A3 & 0.01 & 9 & 1.0 & 10 & 5.0 && $1\times10^{-6}$ & \e{1} & 11 (10.3) & 1
 && 300 & 0.15 -- 2.34 & \oops{7 -- 3}\\ 
A4 & 0.01 & 9 & \e{0.1} & 10 & 5.0 && $1\times10^{-6}$ & 2 & 11 (10.3) & 1
 && 300 & 0.07 -- 2.34 & \oops{8 -- 3}\\ 
A5 & \e{0.09} & 9 & \e{0.1} & 10 & 5.0 && $1\times10^{-6}$ & 2 & 11 (10.3) & \e{3}
 && 300 & 0.07 -- 2.34 & \oops{8 -- 3}\\ 
A5-1 & 0.01 & 9 & 1.0 & 10 & 5.0 && $1\times10^{-6}$ & 2 & 11 (10.3) & \e{3}
 && 300 & 0.15 -- 2.34 & \oops{7 -- 3}\\ 
A5-2 & \e{0.09} & 9 & 1.0 & 10 & 5.0 && $1\times10^{-6}$ & 2 & 11 (10.3) & 1
 && 300 & 0.15 -- 2.34 & \oops{7 -- 3}\\ 
A6 & 0.01 & 9 & 1.0 & \e{20} & 5.0 && $1\times10^{-6}$ & 2 & 11 (10.3) & 1
 && 600 & 0.59 -- 4.69 & \oops{6 -- 3}\\ 

\hline\sidehead{Supersonic Wind Branch ($M_\ast^{\rm eff}=1$~\Msun)}
\hline\\[-2ex]
SFf & 0.10 & 10 & 1.0 & 20 &    10.0 && $1\times10^{-6}$ & 10 & 10 (9.6) & 3
 && 1000 & 0.98 -- 7.81 & \oops{6 -- 3}\\ 
SFm & 0.10 & 10 & 1.0 & 20 & \e{5.0} && $1\times10^{-6}$ & 10 & 10 (9.6) & 3
 && 1000 & 0.98 -- 7.81 & \oops{6 -- 3}\\ 
SFs & 0.10 & 10 & 1.0 & 20 & \e{2.2} && $1\times10^{-6}$ & 10 & 10 (9.6) & 3
 && 1000 & 0.98 -- 7.81 & \oops{6 -- 3}\\ 
SFx & 0.10 & 10 & 1.0 & 20 & \e{0.5} && $1\times10^{-6}$ & 10 & 10 (9.6) & 3
 && 1000 & 0.98 -- 7.81 & \oops{6 -- 3}\\ 
SSf & 0.10 & 10 & 1.0 & 20 &    10.0 && $1\times10^{-6}$ & 10 &  5 (4.2) & 3
 && \oops{3}000 & 0.73 -- 11.7 & \oops{8 -- 4}\\ 
SSm & 0.10 & 10 & 1.0 & 20 & \e{5.0} && $1\times10^{-6}$ & 10 &  5 (4.2) & 3
 && \oops{3}000 & 0.73 -- 11.7 & \oops{8 -- 4}\\ 
SSs & 0.10 & 10 & 1.0 & 20 & \e{2.2} && $1\times10^{-6}$ & 10 &  5 (4.2) & 3
 && \oops{3}000 & 0.73 -- 11.7 & \oops{8 -- 4}\\ 
SSx & 0.10 & 10 & 1.0 & 20 & \e{0.5} && $1\times10^{-6}$ & 10 &  5 (4.2) & 3
 && \oops{3}000 & 0.73 -- 11.7 & \oops{8 -- 4}\\ 

\hline\sidehead{Transonic Wind Branch ($M_\ast^{\rm eff}=1$~\Msun)}
\hline\\[-2ex]
T & 0.25 & 25 & 1.0 & 100 & 10.0 && $5\times10^{-5}$ & 20 & 3.4 (1.7) &  3
 && 2000 & 0.98 -- 15.6 & \oops{7 -- 3} 
\enddata
\tablecomments{Each column represents 
[1] model name, 
[2] perturbing object's mass $M_p$, 
[3] Bondi accretion radius $r_B=GM_p/\cs^2$, 
[4] gravitational softening radius $r_s$, 
[5] orbital radius $r_p$, 
[6] orbital Mach number $\mach_p$, 
[7] stellar mass loss rate $\dot{M}_\ast$, 
[8] stellar size $r_\ast$ that generates the wind, 
[9] wind Mach number $\mach_w$ on average over the simulation domain (and at $r_p$), 
[10] sound speed of background gas $\cs$, 
[11] \oops{half of domain size $L$}, 
[12] \oops{range of\ } spatial resolution $\Delta \oops{L}$\oops{,
[13] corresponding range of refinement level with the mother grids of
64$\times$64$\times$32 in $(x,y,z)$-directions}.
Mass and gravitational softening radius of the central star are 1\Msun\ 
and 1~AU, respectively. Underlines emphasize the main difference of 
the respective models from the fiducial models appearing at the top 
in categories ``A'', ``SF'', and ``SS'', respectively.}
\end{deluxetable}

\begin{thebibliography}{}
\bibitem[Aikawa \& Herbst(1999)]{aik99} Aikawa, Y., \& Herbst, E.\ 1999, \aap, 351, 233
\bibitem[Binney \& Tremaine(2008)]{bin08} Binney, J., \& Tremaine, S.\ 2008, Galactic Dynamics (Princeton, NJ: Princeton Univ. Press)
\bibitem[Dinh-V.-Trung \& Lim(2009)]{tru09} Dinh-V.-Trung, \& Lim, J.\ 2009, \apj, 701, 292 
\bibitem[Edgar(2004)]{edg04} Edgar, R.\ 2004, \nar, 48, 843
\bibitem[Edgar et al.(2008)]{edg08} Edgar, R.~G., Nordhaus, J., Blackman, E.~G., \& Frank, A.\ 2008, \apjl, 675, L101
\bibitem[Fong et al.(2006)]{fon06} Fong, D., Meixner, M., Sutton, E.~C., Zalucha, A., \& Welch, W.~J.\ 2006, \apj, 652, 1626 
\bibitem[Fryxell et al.(2000)]{fry00} Fryxell, B,. Olson, K., Ricker, P., Timmes, F.\ X., Zingale, M., Lamb, D.\ Q., MacNeice, P., Rosner, R., Truran, J.\ W., \& Tufo, H.\ 2000, \apjs, 131, 273
\bibitem[Fukagawa et al.(2004)]{fuk04} Fukagawa, M., et al.\ 2004, \apjl, 605, L53 
\bibitem[Gilman(1972)]{gil72} Gilman, R.~C.\ 1972, \apj, 178, 423 
\bibitem[Goldreich \& Tremaine(1979)]{gol79} Goldreich, P., \& Tremaine, S.\ 1979, \apj, 233, 857 
\bibitem[Habing \& Olofsson(2003)]{hab03} Habing, H.~J., \& Olofsson, H.\ 2003, Asymptotic Giant Branch Stars (New York: Springer)
\bibitem[Hayashi(1981)]{hay81} Hayashi, C.\ 1981, Progress of Theoretical Physics Supplement, 70, 35 
\bibitem[He(2007)]{he07} He, J.~H.\ 2007, \aap, 467, 1081
\bibitem[Hrivnak et al.(2001)]{hri01} Hrivnak, B.~J., Kwok, S., \& Su, K.~Y.~L.\ 2001, \aj, 121, 2775
\bibitem[Hurley et al.(2000)]{hur00} Hurley, J.~R., Pols, O.~R., \& Tout, C.~A.\ 2000, \mnras, 315, 543 
\bibitem[Kim(2011)]{kim11} Kim, H.\ 2011, \apj, 739, 102 (\paper)
\bibitem[Kim \& Kim(2007)]{kim07} Kim, H., \& Kim, W.-T.\ 2007, \apj, 665, 432
\bibitem[Kim \& Kim(2009)]{kim09} Kim, H., \& Kim, W.-T.\ 2009, \apj, 703, 1278
\bibitem[Kim et al.(2008)]{kim08} Kim, H., Kim, W.-T., \& S{\'a}nchez-Salcedo, F.~J.\ 2008, \apjl, 679, L33
\bibitem[Kim(2010)]{kwt10} Kim, W.-T.\ 2010, \apj, 725, 1069
\bibitem[Kwok et al.(2001)]{kwo01} Kwok, S., Su, K.~Y.~L., \& Stoesz, J.~A.\ 2001, Astrophysics and Space Science Library, 265, 115
\bibitem[Lafon \& Berruyer(1991)]{laf91} Lafon, J.-P.~J., \& Berruyer, N.\ 1991, \aapr, 2, 249
\bibitem[Lamers \& Cassinelli(1999)]{lam99} Lamers, H.~J.~G.~L.~M., \& Cassinelli, J.~P.\ 1999, Introduction to Stellar Winds (Cambridge, UK: Cambridge Univ. Press)
\bibitem[Lin et al.(2006)]{lin06} Lin, S.-Y., Ohashi, N., Lim, J., Ho, P.~T.~P., Fukagawa, M., \& Tamura, M.\ 2006, \apj, 645, 1297
\bibitem[MacNeice et al.(1999)]{mac99} MacNeice, P., Olson, K.~M., Mobarry, C., de Fainchtein, R., \& Packer, C.\ 1999, CPC, 126, 3
\bibitem[Masset(2008)]{mas08} Masset, F.~S.\ 2008, EAS Publications Series, 29, 165 
\bibitem[Mastrodemos \& Morris(1999)]{mas99} Mastrodemos, N., \& Morris, M.\ 1999, \apj, 523, 357 
\bibitem[Mauron \& Huggins(1999)]{mau99} Mauron, N., \& Huggins, P.~J.\ 1999, \aap, 349, 203
\bibitem[Mauron \& Huggins(2006)]{mau06} Mauron, N., \& Huggins, P.~J.\ 2006, \aap, 452, 257 
\bibitem[Morris et al.(2006)]{mor06} Morris, M., Sahai, R., Matthews, K., Cheng, J., Lu, J., Claussen, M., \& S{\'a}nchez-Contreras, C.\ 2006, Planetary Nebulae in our Galaxy and Beyond, 234, 469 
\bibitem[Namouni(2010)]{nam10} Namouni, F.\ 2010, \mnras, 401, 319 
\bibitem[Ostriker(1999)]{ost99} Ostriker, E.~C.\ 1999, \apj, 513, 252
\bibitem[Parker(1958)]{par58} Parker, E.~N.\ 1958, \apj, 128, 664
\bibitem[Sahai et al.(1998)]{sah98} Sahai, R., et al.\ 1998, \apj, 493, 301 
\bibitem[Simis et al.(2001)]{sim01} Simis, Y.~J.~W., Icke, V., \& Dominik, C.\ 2001, \aap, 371, 205 
\bibitem[Soker(1994)]{sok94} Soker, N.\ 1994, \mnras, 270, 774 
\bibitem[Villaver \& Livio(2009)]{vil09} Villaver, E., \& Livio, M.\ 2009, \apjl, 705, L81
\bibitem[Willson \& Hill(1979)]{wil79} Willson, L.~A., \& Hill, S.~J.\ 1979, \apj, 228, 854
\bibitem[Winters et al.(2000)]{win00} Winters, J.~M., Le Bertre, T., Jeong, K.~S., Helling, C., \& Sedlmayr, E.\ 2000, \aap, 361, 641
\bibitem[Wood(1979)]{woo79} Wood, P.~R.\ 1979, \apj, 227, 220 
\end{thebibliography}
\end{document}